\documentclass[aps,prd,preprint,nofootinbib,superscriptaddress]{revtex4-2}

\usepackage[a4paper,top=2.5cm,bottom=2.5cm,left=3.0cm,right=2.5cm]{geometry}
\usepackage{amsmath,amssymb,amsfonts}
\usepackage{bm}
\usepackage{graphicx}
\usepackage{booktabs}
\usepackage{hyperref}
\usepackage{xcolor}
\usepackage{multirow}

\begin{document}
	
	%======================================================================
	\title{
	 Color‑octet contributions via two photons for double‑$J\psi$ production at B factories
	}
	%======================================================================

	\author{Meng-Kun Jia $^{(a)}$}
	\author{Yi-Jie Li $^{(a)}$}
	\author{Guang-Zhi Xu $^{(a)}$}
	\email{xuguangzhi@lnu.edu.cn}
	\author{Kui-Yong Liu $^{(b,a)}$}
	\email{liukuiyong@lnu.edu.cn}
	\affiliation{ {\footnotesize (a)~School of Physics, Liaoning University, Shenyang 110036, China}\\
		{\footnotesize (b)~School of Physics and Electronic Technology, Liaoning Normal University, Dalian 116029, China}}

	%======================================================================
\begin{abstract}
   We calculate the two-photon color-octet contributions to double $J/\psi$ production in $e^+e^-$ annihilation at $\mathcal O(\alpha^4\alpha_s)$ for the ${}^3S_1^{[1]}+{}^1S_0^{[8]}+g$ and ${}^3S_1^{[1]}+{}^3P_J^{[8]}+g$ channels. Using the Chao et al.\ long-distance matrix element (LDME) set, the summed CO cross section is $-0.0531\mathrm{fb}$ at $\sqrt{s}=10.6\mathrm{GeV}$, corresponding to $-13\%$ of the next-to-leading-order color-singlet result and approximately $1.7$ times the previously calculated summed single-photon CO contribution. Across representative LDME sets, the total CO contribution spans $-0.397$ to $0.201~\mathrm{fb}$. These large variations highlight the need for precise future experimental measurements to constrain the color-octet LDMEs.
\end{abstract}

\maketitle

%================================================================
\section{Introduction}
\label{sec:introduction}
%================================================================

Heavy quarkonium production provides an important test of nonrelativistic QCD (NRQCD) factorization, in which perturbative short distance coefficients (SDCs) are combined with long-distance matrix elements (LDMEs) describing quarkonium formation \cite{Bodwin:1994jh}. In addition to color-singlet (CS) states, color-octet (CO) configurations contribute through higher Fock components. Their numerical importance depends on both the relevant LDMEs and the short-distance production process, and kinematic or fragmentation enhancements can partly compensate the LDME suppression in specific reactions or phase space regions \cite{Braaten:1995CO}. The sizable spread among existing $J/\psi$ CO LDME determinations \cite{Butenschoen:2011yh,Butenschoen:2012px,Chao:2012iv,Gong:2012ug,Bodwin:2014gia} therefore motivates studying CO contributions in different production processes. Color-octet contributions to inclusive $J/\psi$ production at $B$ factories have also been studied in processes involving initial-state photon radiation, at higher orders in QCD, and with soft-gluon effects taken into account ~\cite{Wang:2003fw,Zhang:2009CO,Chen:2022SGF}.

Electron--positron annihilation at $B$-factory energies provides a clean environment for such studies. In double-$J/\psi$ production, the leading CS process, $e^+e^-\to J/\psi+J/\psi$, proceeds through two virtual photons and starts at $\mathcal O(\alpha^4)$. Despite the additional electromagnetic suppression, the cross section is substantially enhanced by photon-fragmentation diagrams at $B$-factory energies \cite{Bodwin:2002JpsiJpsi,Bodwin:2002hg,Bodwin:2006ke}. QCD and relativistic corrections have been investigated \cite{Gong:2008ce,Fan:2012dy}, and the CS contribution has recently been calculated through $\mathcal O(\alpha_s^2)$ \cite{Sang:2023double,Huang:2024double}. 
Recent NNLO QCD calculations at $B$ factories have also considered $e^+e^-\to J/\psi+\chi_{cJ}$ \cite{Liu:2026ray} and direct color-singlet $J/\psi$ production in association with with open charm \cite{Chang:2026vqf}.
Double charmonium production has also been studied experimentally through the recoil mass against a reconstructed $J/\psi$ \cite{Belle:2004ab,BaBar:2005de}.

Previous studies of CO effects in double-$J/\psi$ production at $B$ factories focused mainly on the single-photon contributions through $\mathcal O(\alpha^2\alpha_s^3)$ \cite{Feng:2017}. The two-photon ${}^3S_1^{[1]}+{}^1S_0^{[8]}+g$ and ${}^3S_1^{[1]}+{}^3P_J^{[8]}+g$ channels, which first arise at $\mathcal O(\alpha^4\alpha_s)$, were not included. These channels constitute a distinct two-photon CO contribution and are of particular interest because of the photon-fragmentation enhancement in double-$J/\psi$ production. The ${}^1S_0^{[8]}$ contribution is infrared finite, whereas the soft singularity in the ${}^3P_J^{[8]}$ channel is canceled through the renormalization of the corresponding NRQCD matrix element \cite{Petrelli:1997ge}. We isolate the soft region using a phase-space soft cutoff following the phase-space-slicing approach of Ref.~\cite{Harris:2001sx}.

In this work, we calculate these two-photon CO contributions and study their integrated cross sections, center of mass energy dependence, and the scaled-energy distribution of the produced $J/\psi$. We also examine the dependence on the charm-quark mass, the NRQCD factorization scale, and representative $J/\psi$ LDME sets. Particular attention is paid to the interplay between the ${}^1S_0^{[8]}$ and ${}^3P_J^{[8]}$ contributions and to the comparison with the corresponding two-photon CS and previously calculated single-photon CO contributions.

%======================================================================
\section{Theoretical framework}
\label{sec:theory}
%======================================================================

\begin{figure}[t]
	\centering
	\includegraphics[width=0.75\linewidth]{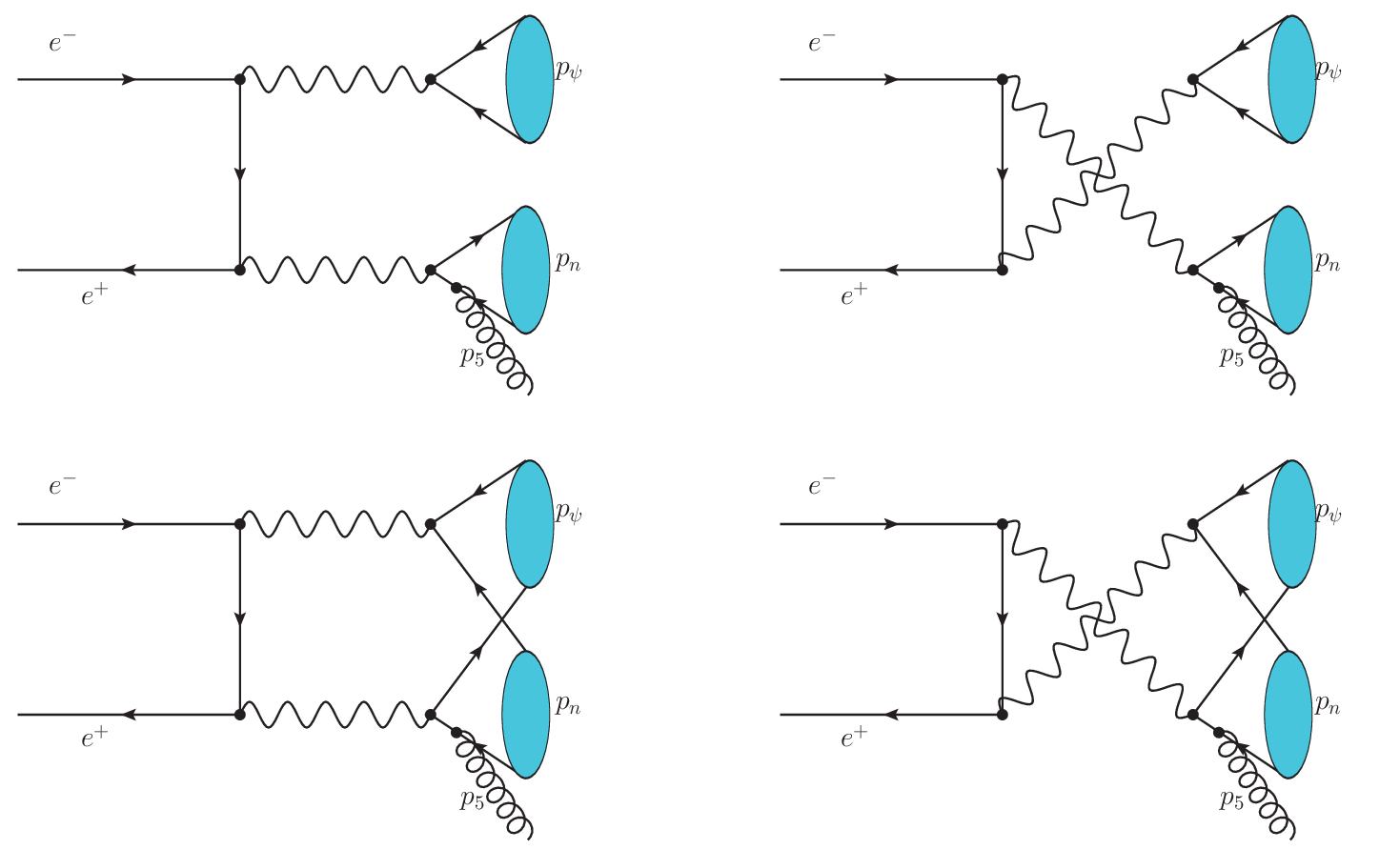}
	\centering
\caption{
	Representative tree-level Feynman diagrams for the two-photon
	${}^3S_1^{[1]}+{}^1S_0^{[8]}+g$ and
	${}^3S_1^{[1]}+{}^3P_J^{[8]}+g$ processes.
	Four representative diagrams are shown; the remaining four are
	obtained by reversing the fermion flow.
}
	\label{fig:feynman}
\end{figure}

The quark level CO subprocesses via two-photon exchange, which arise at $\mathcal O(\alpha^4\alpha_s)$, read
\begin{align}
	e^-(l_1)+e^+(l_2)
	&\to
	c\bar c({}^3S_1^{[1]})(p_\psi)
	+c\bar c({}^1S_0^{[8]})(p_n)+g(p_5),
	\nonumber\\
	e^-(l_1)+e^+(l_2)
	&\to
	c\bar c({}^3S_1^{[1]})(p_\psi)
	+c\bar c({}^3P_J^{[8]})(p_n)+g(p_5),
	\qquad J=0,1,2 .
	\label{eq:processes}
\end{align}
For each channel, there are eight tree-level Feynman diagrams. Four representative diagrams are shown in Fig.~\ref{fig:feynman}, while the remaining four are obtained by reversing the fermion flow. 
Electromagnetic gauge invariance is verified through the Ward identities for both virtual photons.

At leading order in the heavy-quark velocity expansion, the invariant mass of each intermediate $c\bar c$ pair is taken to be $M=2m_c$, such that $p_\psi^2=p_n^2=M^2$. 
Within the NRQCD framework, heavy-quark spin symmetry gives the relation
\begin{equation}
	\langle{\cal O}^{J/\psi}({}^3P_J^{[8]})\rangle
	=
	(2J+1)
	\langle{\cal O}^{J/\psi}({}^3P_0^{[8]})\rangle ,
	\label{eq:hqss}
\end{equation}
With this relation, the two-photon CO cross section is factorizated as
\begin{align}
    \sigma^{\rm CO}
	= \sigma_{{}^1S_0^{[8]}} + \sigma_{{}^3P_J^{[8]}} =
	\langle{\cal O}^{J/\psi}({}^3S_1^{[1]})\rangle
	\Bigg[
	\hat\sigma_S
	\langle{\cal O}^{J/\psi}({}^1S_0^{[8]})\rangle
	+
	\hat\sigma_{P}
	\langle{\cal O}^{J/\psi}({}^3P_0^{[8]})\rangle
	\Bigg].
	\label{eq:nrqcd_factorization}
\end{align}
Here $\hat\sigma_S$ and $\hat\sigma_{P}$ denote the corresponding ${}^1S_0^{[8]}$ and ${}^3P_J^{[8]}$ short distance coefficients.

The ${}^1S_0^{[8]}$ contribution is infrared finite over the full three-body phase space. In contrast, the ${}^3P_J^{[8]}$ contribution contains a soft singularity when the emitted-gluon energy approaches zero. 
Following the phase space slicing method~\cite{Harris:2001sx}, a soft cutoff $\delta_s$ on the final state gluon energy is introduced to separate the phase space according to
\begin{equation}
	E_5<\frac{\delta_s\sqrt{s}}{2}
	\quad {\rm (soft)},\qquad
	E_5\geq\frac{\delta_s\sqrt{s}}{2}
	\quad {\rm (hard)} .
	\label{eq:slicing}
\end{equation}
The hard part is finite and is evaluated in four dimensions, whereas the soft part is treated analytically in $D=4-2\epsilon$ dimensions. 
The soft short-distance coefficient $\hat\sigma_{P}^{\rm s}$ is expressed as the leading-order CS short distance coefficient (with $c\bar c({}^3S_1^{[1]})+ c\bar c({}^3S_1^{[1]})$ final states) multiplied by a soft factor $\mu_\epsilon^s$,
\begin{equation}
	\hat\sigma_{P}^{\rm s}
	=
	\mu_\epsilon^s\hat\sigma^{\rm CS}_{\rm LO},
	\label{eq:soft_factorization}
\end{equation}
where
\begin{equation}
	\mu_\epsilon^s
	=
	-\frac{2\alpha_s}{\pi m_c^2}
	\left[
	\frac{1}{\epsilon_{\rm IR}}
	+\frac{1}{\beta}
	\ln\frac{1+\beta}{1-\beta}
	+\ln\frac{4\pi\mu_r^2}{s\delta_s^2}
	-\gamma_E-\frac13
	\right],
	\label{eq:mu_soft}
\end{equation}
with
\begin{equation}
	\beta=\sqrt{1-\frac{16m_c^2}{s}} .
	\label{eq:beta}
\end{equation}

The infrared divergence cancels against the NLO term of $\langle{\cal O}^{J/\psi}({}^3S_1^{[1]})\rangle$. The renormalized coefficient multiplying $\langle{\cal O}^{J/\psi}({}^3P_0^{[8]})\rangle$ reads
\begin{equation}
	\mu_\epsilon^c
	=
	-\frac{2\alpha_s}{\pi m_c^2}
	\left[
	\frac{1}{\epsilon_{\rm IR}}
	-\gamma_E+\frac53
	+\ln\frac{\pi\mu_r^2}{\mu_\Lambda^2}
	\right],
	\label{eq:mu_counter}
\end{equation}
where $\mu_\Lambda$ is the NRQCD factorization scale.

Through the matching procedure, the soft shor distance coefficient becomes finite, with the factor given by
\begin{align}
	\mu^{\rm fin}
	=
	\mu_\epsilon^s-\mu_\epsilon^c=
	-\frac{2\alpha_s}{\pi m_c^2}
	\left[
	\frac{1}{\beta}
	\ln\frac{1+\beta}{1-\beta}
	+\ln\frac{4\mu_\Lambda^2}{s\delta_s^2}
	-2
	\right].
	\label{eq:mu_finite}
\end{align}
The $1/\epsilon_{\rm IR}$ pole and the explicit $\mu_r$ dependence therefore cancel. The ${}^3P_J^{[8]}$ short distance coefficient is then
\begin{equation}
	\hat\sigma_P^{\rm fin}
	=
	\hat\sigma_{P}^{{\rm h}}(\delta_s)
	+\mu^{\rm fin}\hat\sigma^{\rm CS}_{\rm LO}.
	\label{eq:pwave_finite}
\end{equation}
The dependence on the cutoff $\delta_s$ cancels between the hard and finite soft contributions in the summed result.

%======================================================================
\section{Numerical results}
\label{sec:results}
%======================================================================

For the numerical calculation, we take $m_c=1.5~{\rm GeV}$, $\alpha=1/137$, $\alpha_s(2m_c)=0.26$ and $\mu_\Lambda=m_c$. 
The strong coupling at renormalization scale $\mu_r=2m_c$ is evaluated at two loops. 
For the baseline prediction, we use the LDME set from Chao et al.~\cite{Chao:2012iv},
\begin{equation}
	\begin{aligned}
		\langle{\cal O}^{J/\psi}({}^3S_1^{[1]})\rangle
		&=1.16~{\rm GeV}^3,
		&
		\langle{\cal O}^{J/\psi}({}^1S_0^{[8]})\rangle
		&=8.9\times10^{-2}~{\rm GeV}^3,
		\\
		\langle{\cal O}^{J/\psi}({}^3S_1^{[8]})\rangle
		&=3.0\times10^{-3}~{\rm GeV}^3,
		&
		\frac{
			\langle{\cal O}^{J/\psi}({}^3P_0^{[8]})\rangle
		}{m_c^2}
		&=5.6\times10^{-3}~{\rm GeV}^3 .
	\end{aligned}
	\label{eq:central_ldmes}
\end{equation}

\begin{figure}[t]
	\centering
	\includegraphics[width=0.75\linewidth]{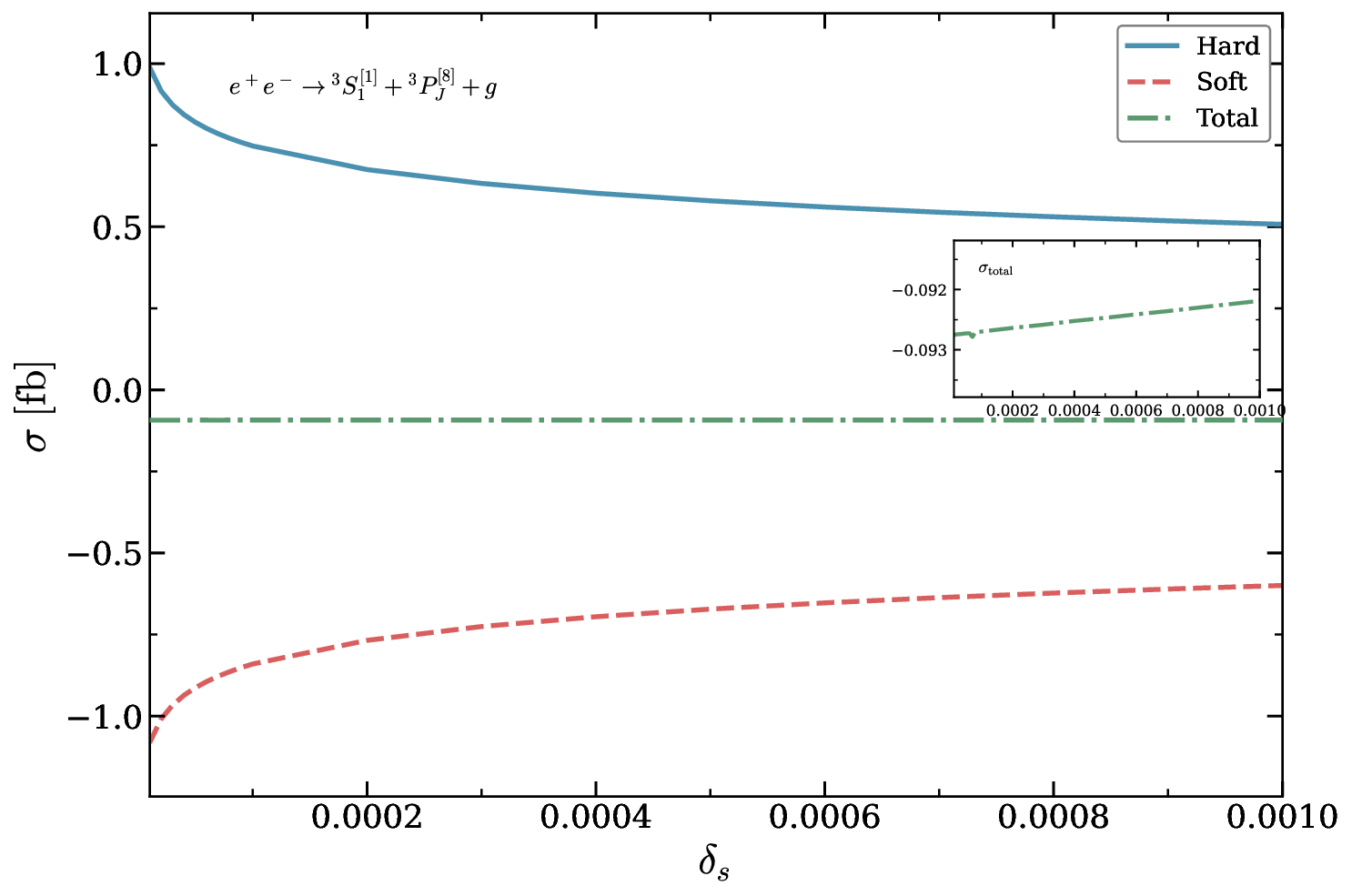}
	\caption{
		Dependence of the ${}^3P_J^{[8]}$ contribution on the soft cutoff $\delta_s$. The inset shows the residual variation of the sum.
	}
	\label{fig:delta}
\end{figure}

As a numerical check of the soft treatment, Fig.~\ref{fig:delta} shows the dependence of the ${}^3P_J^{[8]}$ contribution on the soft cutoff $\delta_s$. The hard and soft contributions vary separately with $\delta_s$, whereas their sum becomes stable for sufficiently small values of the cutoff. We therefore use $\delta_s=10^{-4}$ in the numerical calculation.

At $\sqrt{s}=10.6~{\rm GeV}$, we obtain
\begin{align}
	\sigma_{{}^1S_0^{[8]}}
	&=
	0.0396~{\rm fb},
	\nonumber\\
	\sigma_{{}^3P_J^{[8]}}
	&=
-0.0927~{\rm fb},
	\nonumber\\
	\sigma^{\rm CO}
	&=
-0.05311~{\rm fb}.
	\label{eq:central_results}
\end{align}

Table~\ref{tab:scale_dependence} summarizes the dependence of the CO cross sections on $m_c$, $\mu_r$, and $\mu_\Lambda$. The hard and soft contributions to the ${}^3P_J^{[8]}$ channel are listed separately. The results show that the ${}^3P_J^{[8]}$ contribution is sensitive to $\mu_\Lambda$, whereas the ${}^1S_0^{[8]}$ contribution remains unchanged when $\mu_\Lambda$ is varied.

\begin{table*}[t]
	\centering
	\caption{
		Dependence of the color octet cross sections on
		the charm-quark mass $m_c$, the renormalization
		scale $\mu_r$, and the NRQCD factorization scale
		$\mu_\Lambda$ at $\sqrt{s}=10.6~\mathrm{GeV}$.
		The hard and soft contributions to the
		${}^3P_J^{[8]}$ channel are shown separately.
		All cross sections are given in fb.
	}
	\label{tab:scale_dependence}
	\scriptsize
	\setlength{\tabcolsep}{4pt}
	\renewcommand{\arraystretch}{1.15}
	\begin{tabular}{ccccccccc}
		\toprule
		$m_c$ (GeV)
		& $\mu_r$
		& $\mu_\Lambda$
		& $\alpha_s(\mu_r)$
		& $\sigma_P^{\rm hard}$
		& $\sigma_P^{\rm soft}$
		& $\sigma_P$
		& $\sigma_S$
		& $\sigma_{\rm CO}$
		\\
		\midrule
		
		% mc = 1.4 GeV
		1.4 & $m_c$ & $m_c$
		& 0.386 & 1.193 & -1.338
		& -0.145 & 0.068 & -0.077 \\
		
		1.4 & $2m_c$ & $m_c$
		& 0.267 & 0.825 & -0.926
		& -0.100 & 0.047 & -0.053 \\
		
		1.4 & $\sqrt{s}/2$ & $m_c$
		& 0.211 & 0.652 & -0.731
		& -0.079 & 0.037 & -0.042 \\
		
		\cmidrule(lr){2-9}
		
		1.4 & $2m_c$ & $m_c/2$
		& 0.267 & 0.825 & -0.846
		& -0.020 & 0.047 & 0.027 \\
		
		1.4 & $2m_c$ & $2m_c$
		& 0.267 & 0.825 & -1.006
		& -0.180 & 0.047 & -0.133 \\
		
		\midrule
		
		% mc = 1.5 GeV
		1.5 & $m_c$ & $m_c$
		& 0.369 & 1.061 & -1.193
		& -0.132 & 0.056 & -0.075 \\
		
		1.5 & $2m_c$ & $m_c$
		& 0.260 & 0.748 & -0.840
		& -0.093 & 0.040 & -0.053 \\
		
		1.5 & $\sqrt{s}/2$ & $m_c$
		& 0.211 & 0.607 & -0.682
		& -0.075 & 0.032 & -0.043 \\
		
		\cmidrule(lr){2-9}
		
		1.5 & $2m_c$ & $m_c/2$
		& 0.260 & 0.748 & -0.768
		& -0.020 & 0.040 & 0.019 \\
		
		1.5 & $2m_c$ & $2m_c$
		& 0.260 & 0.748 & -0.913
		& -0.165 & 0.040 & -0.126 \\
		
		\midrule
		
		% mc = 1.6 GeV
		1.6 & $m_c$ & $m_c$
		& 0.354 & 0.946 & -1.066
		& -0.120 & 0.046 & -0.074 \\
		
		1.6 & $2m_c$ & $m_c$
		& 0.253 & 0.676 & -0.762
		& -0.086 & 0.033 & -0.053 \\
		
		1.6 & $\sqrt{s}/2$ & $m_c$
		& 0.211 & 0.564 & -0.635
		& -0.071 & 0.027 & -0.044 \\
		
		\cmidrule(lr){2-9}
		
		1.6 & $2m_c$ & $m_c/2$
		& 0.253 & 0.676 & -0.696
		& -0.020 & 0.033 & 0.013 \\
		
		1.6 & $2m_c$ & $2m_c$
		& 0.253 & 0.676 & -0.827
		& -0.151 & 0.033 & -0.119 \\
		
		\bottomrule
	\end{tabular}
\end{table*}

For the adopted baseline LDME set, the ${}^1S_0^{[8]}$ and ${}^3P_J^{[8]}$ contributions have opposite signs, with the latter being larger in magnitude. Their partial cancellation therefore substantially reduces the summed two-photon CO contribution. The charm quark mass dependence is relatively mild, whereas the residual NRQCD factorization scale dependence is sizable and can even change the sign of the summed result.

The two-photon CS cross section up to next to leading order in $\alpha_s$, evaluated with the same charm-quark mass and CS LDME, is $0.42~\mathrm{fb}$~\cite{Gong:2008ce}. The baseline two-photon CO contribution therefore amounts to approximately $-13\%$ of the CS result.
For comparison, the summed single photon CO contribution obtained with the same LDME set from Chao et al.~\cite{Chao:2012iv} is about  $0.032~{\rm fb}$~\cite{Feng:2017}. The magnitude of the present two-photon CO contribution is thus about $1.7$ times this value.

\begin{figure}[t]
	\centering
	\includegraphics[width=0.75\linewidth]{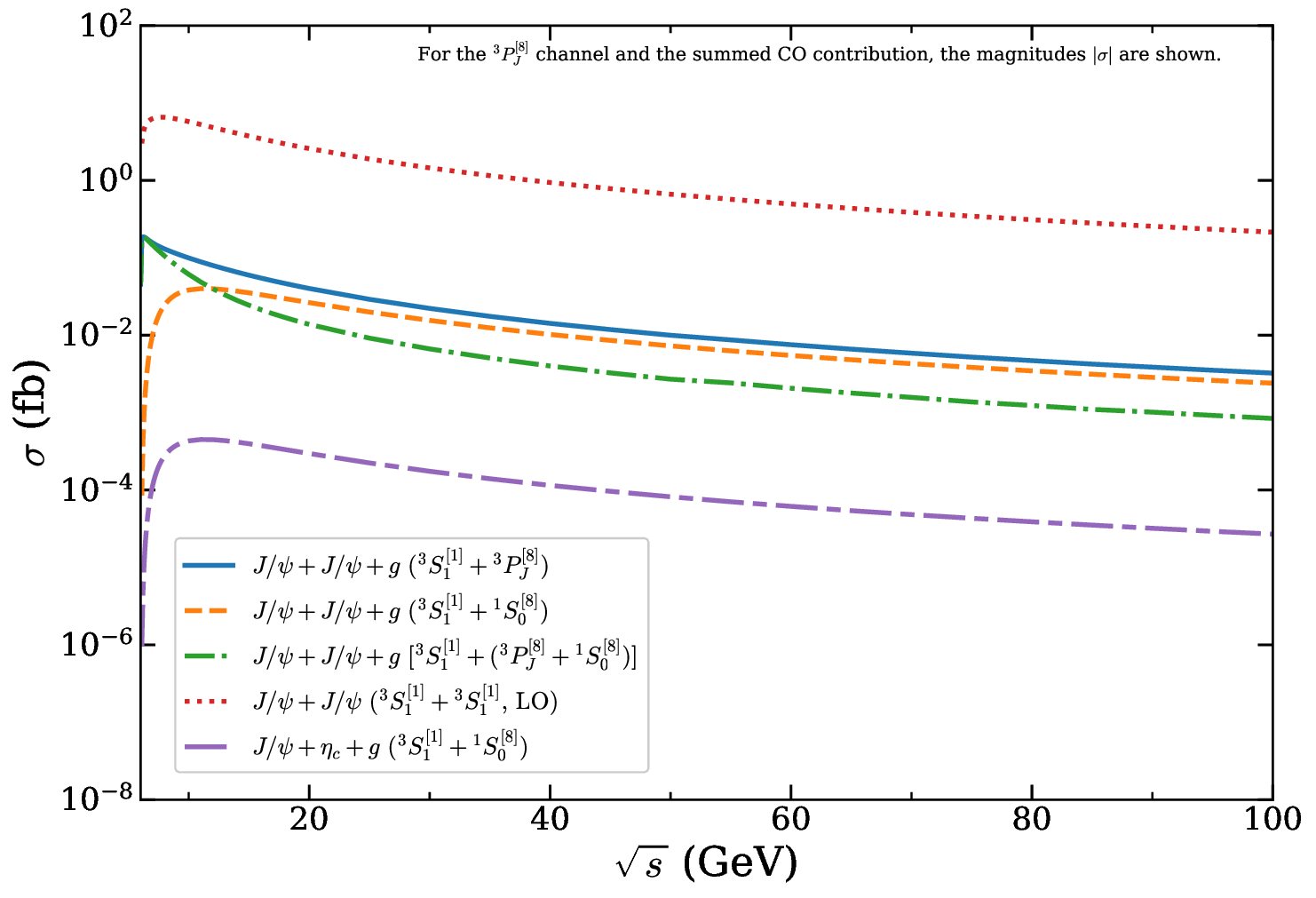}
	\caption{
		Energy dependence of the two-photon CS and CO contributions to double $J/\psi$ production. The two-photon ${}^3S_1^{[1]}+{}^1S_0^{[8]}+g$ contribution to $J/\psi+\eta_c+X$ is also shown for comparison. The magnitudes of the renormalized ${}^3P_J^{[8]}$ and summed CO contributions are displayed.
	}
	\label{fig:energy_dependence}
\end{figure}

We next examine the energy dependence of the two-photon contributions, as shown in Fig.~\ref{fig:energy_dependence}. 
The leading order CS curve is also included for comparison.
Above the peak region, the two-photon CS and CO contributions exhibit similar high energy falloffs. To quantify this behavior, we define the local effective power $n_{\rm eff}=-d\ln|\sigma|/d\ln s$. In the range $\sqrt{s}=20$--$100~{\rm GeV}$, $n_{\rm eff}$ is approximately $0.8$ for the CS contribution as well as for both the ${}^1S_0^{[8]}$ and ${}^3P_J^{[8]}$ contributions. The corresponding cross sections therefore decrease approximately as $s^{-0.8}$ over this energy range, indicating a gradual tendency toward a $1/s$-type behavior at higher energies. Such a slow falloff is consistent with the leading-power enhancement associated with photon-fragmentation diagrams ~\cite{Bodwin:2002JpsiJpsi,Bodwin:2002hg,Bodwin:2006ke}.

For comparison, we also show the two-photon ${}^3S_1^{[1]}+{}^1S_0^{[8]}+g$ contribution to $J/\psi+\eta_c+X$ in the figure. 
The values of $\eta_c$ CO LDMEs are obtained using heavy-quark spin symmetry (HQSS)~\cite{Bodwin:1994jh} as
$
	\langle {\cal O}^{\eta_c}({}^1S_0^{[8]})\rangle
	=
	\frac{1}{3}
	\langle {\cal O}^{J/\psi}({}^3S_1^{[8]})\rangle
	=
	1.0\times10^{-3}~{\rm GeV}^3,
$
where $\langle {\cal O}^{J/\psi}({}^3S_1^{[8]})\rangle$ is also taken from Ref.~\cite{Chao:2012iv}.
Its energy dependence is similar to that of the corresponding ${}^1S_0^{[8]}$ contribution in double-$J/\psi$ production, while its overall magnitude is substantially smaller around $B$-factory energies, mainly because of the smaller associated $\eta_c$ CO LDME. The smallness of this particular two-photon contribution does not conflict with the sizable CO effects found in inclusive $J/\psi+\eta_c+X$ production, where additional short distance contributions can be important~\cite{Sun:2026hff}.

It is also worth noting that sizable CO contributions can arise in $\gamma^*/Z^*$-mediated double-quarkonium production at future $Z$ factories, where gluon fragmentation into a ${}^3S_1^{[8]}$ heavy quark pair can play an important role ~\cite{Wang:2026gso}. This represents a different short distance origin from the two-photon enhancement considered here.

For the differential distribution, we define
\begin{align}
	\sigma_1 & =
	\sigma_{{}^3P_J^{[8]}}
	+\sigma^{\rm CS}_{\rm{NLO}(\alpha_s)},
	\\
	\sigma_2 & = \sigma_{{}^1S_0^{[8]}},
\end{align}
and
\begin{equation}
	\sigma_{\rm sum}=\sigma_1+\sigma_2 .
\end{equation}
The same definitions are applied bin by bin to the differential distribution, with all contributions combined with their original signs.

\begin{figure}[t]
	\centering
	\includegraphics[width=0.72\linewidth]{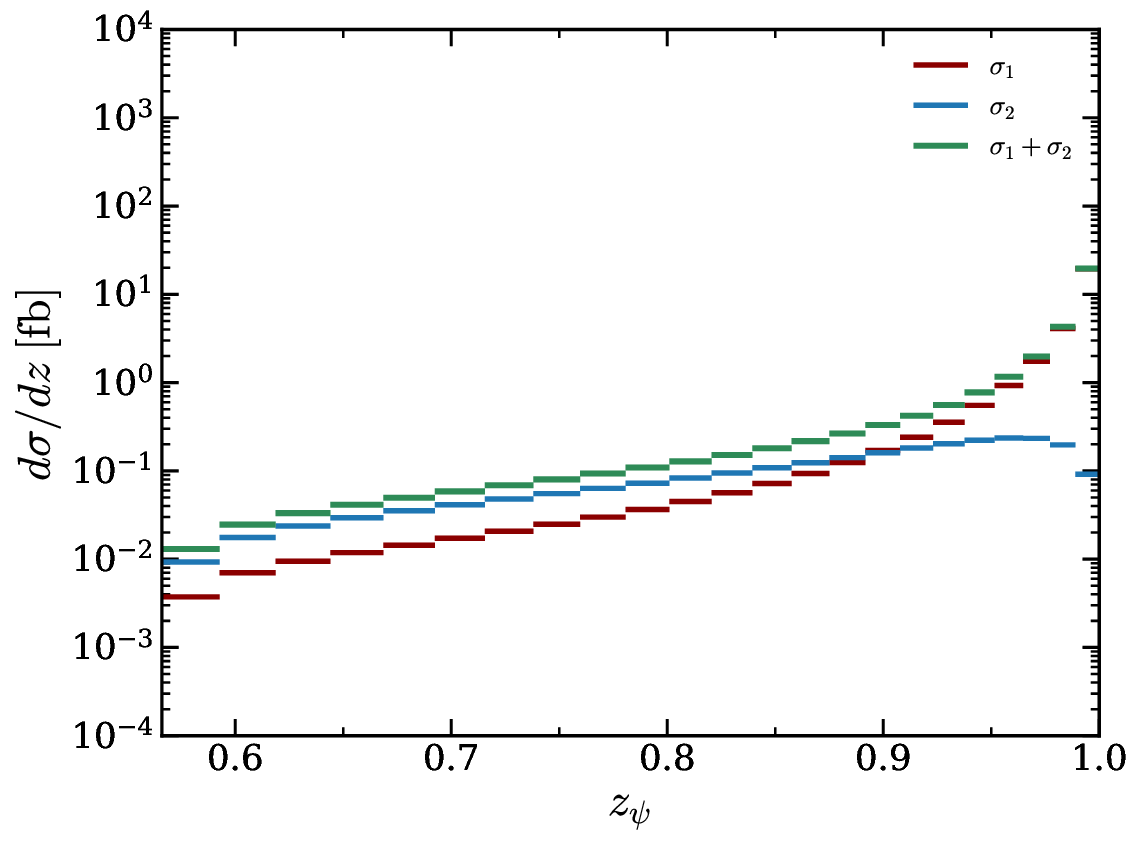}
	\caption{
		Bin-averaged scaled-energy distribution of the produced 	$J/\psi$, with $z_\psi=2E_\psi/\sqrt{s}$. The contributions $\sigma_1$ and $\sigma_2$ are defined in the text, while ``sum'' denotes their signed sum. The two-body ${}^3S_1^{[1]}+{}^3S_1^{[1]}$ contribution contained in $\sigma_1$ is localized at $z_\psi=1$.
	}
	\label{fig:zpsi}
\end{figure}

The scaled energy ($z_\psi=\frac{2E_\psi}{\sqrt{s}}$) distribution of the produced $J/\psi$ is shown in Fig.~\ref{fig:zpsi} as a bin-averaged differential cross section. 
The bins are uniformly spaced in the recoil invariant mass of one $J/\psi$, with a bin width of $0.2~\mathrm{GeV}$, and the resulting distribution is binned in $z_\psi$.
The contribution from ${}^1S_0^{[8]}$ channel extends broadly over the allowed kinematic region, reflecting the regular behavior. 
In contrast, the differential distribution grows toward the upper endpoint, owing to the stronger endpoint dependence of the ${}^3P_J^{[8]}$ contribution. 
Away from the endpoint, the values for the ${}^3P_J^{[8]}$ channel are positive. 
The endpoint bin value obtained by summing the soft and hard contributions near the soft cut is negative and exhibits a discontinuity relative to the hard-region distribution. 
Upon including the next-to-leading-order contribution of the CS channel in $\alpha_s$, which is at the same order as the CO channels, this jump behavior disappears.
We stress that this feature is sensitive to the magnitude and sign of the ${}^3P_0^{[8]}$ LDME.

Finally, we examine the dependence on the $J/\psi$ LDMEs using several representative determinations. The LDME inputs and the corresponding two-photon CO cross sections are summarized in Table~\ref{tab:ldme_cross}.

\begin{table}[t]
	\centering
	\caption{
		Representative $J/\psi$ LDME sets and the corresponding two-photon CO cross sections at $m_c=1.5~{\rm GeV}$ and $\mu_\Lambda=m_c$. The LDMEs and $\langle{\cal O}({}^3P_0^{[8]})\rangle/m_c^2$ are given in ${\rm GeV}^3$, and the cross sections in fb.
	}
	\label{tab:ldme_cross}
	
	\footnotesize
	\setlength{\tabcolsep}{2.8pt}
	\renewcommand{\arraystretch}{1.12}
	
	\begin{tabular}{lcccccc}
		\toprule
		LDME set
		&
		$\langle{\cal O}({}^3S_1^{[1]})\rangle$
		&
		$\langle{\cal O}({}^1S_0^{[8]})\rangle$
		&
		$\langle{\cal O}({}^3P_0^{[8]})\rangle/m_c^2$
		&
		$\sigma_{{}^1S_0^{[8]}}$
		&
		$\sigma_{{}^3P_J^{[8]}}$
		&
		$\sigma^{\rm CO}$
		\\
		\midrule
		
		Zhang et al.\ \cite{Zhang:2014ybe,Sun:2015pia}
		& $0.65$
		& $0.78\times10^{-2}$
		& $2.01\times10^{-2}$
		& $0.002$
		& $-0.186$
		& $-0.185$
		\\
		
		Butensch\"on--Kniehl \cite{Butenschoen:2011yh}
		& $1.32$
		& $3.04\times10^{-2}$
		& $-4.04\times10^{-3}$
		& $0.015$
		& $0.076$
		& $0.092$
		\\
		
		Chao et al.\ (default) \cite{Chao:2012iv}
		& $1.16$
		& $8.9\times10^{-2}$
		& $5.60\times10^{-3}$
		& $0.040$
		& $-0.093$
		& $-0.053$
		\\
		
		Chao et al.\ (Set 2) \cite{Chao:2012iv}
		& $1.16$
		& $0$
		& $2.40\times10^{-2}$
		& $0.000$
		& $-0.397$
		& $-0.397$
		\\
		
		Chao et al.\ (Set 3) \cite{Chao:2012iv}
		& $1.16$
		& $1.10\times10^{-1}$
		& $0$
		& $0.049$
		& $0.000$
		& $0.049$
		\\
		
		Gong et al.\ \cite{Gong:2012ug}
		& $1.16$
		& $9.7\times10^{-2}$
		& $-9.51\times10^{-3}$
		& $0.043$
		& $0.157$
		& $0.201$
		\\
		
		Bodwin et al.\ \cite{Bodwin:2014gia}
		& $1.16^{\,a}$
		& $9.9\times10^{-2}$
		& $4.89\times10^{-3}$
		& $0.044$
		& $-0.081$
		& $-0.037$
		\\
		
		\bottomrule
	\end{tabular}
	
	\vspace{1mm}
	\begin{minipage}{\linewidth}
		\footnotesize
		$^{a}$ The color-singlet LDME is not given in Ref.~\cite{Bodwin:2014gia}; following Ref.~\cite{Feng:2017}, we use $\langle{\cal O}^{J/\psi}({}^3S_1^{[1]})\rangle =1.16~{\rm GeV}^3$.
	\end{minipage}
\end{table}

The variation among the LDME sets is primarily driven by the ${}^3P_0^{[8]}$ LDME, whose magnitude and even sign differ substantially among the available determinations. For the sets considered here, the summed two-photon CO contribution ranges from $-0.397$ to $0.201~{\rm fb}$, %corresponding to approximately $-7.3\%$ to $3.7\%$ of the LO two-photon CS result. 
corresponding to approximately $-95\%$ to $48\%$ of the NLO CS result. 
In view of the LDME constraints from inclusive \(J/\psi\) measurements at the B-factory and near-threshold kinematic region~\cite{Li:2014fya}, the Butensch"on--Kniehl set gives a ratio of $22\%$. 
Both the magnitude and the sign of the summed CO contribution are therefore sensitive to the LDME choice. 
Such large uncertainties highlight the need for future precise experimental constraints on the ${}^3P_0^{[8]}$ LDME.

%======================================================================
\section{Summary}
\label{sec:summary}
%======================================================================

We compute the two-photon color-octet contributions to double $J/\psi$ production in $e^+e^-$ annihilation for the ${}^3S_1^{[1]}+{}^1S_0^{[8]}+g$ and ${}^3S_1^{[1]}+{}^3P_J^{[8]}+g$ channels at $\mathcal O(\alpha^4\alpha_s)$. The ${}^1S_0^{[8]}$ contribution is infrared finite, while the soft singularity of the ${}^3P_J^{[8]}$ channel cancels against the NLO term of the associated $\langle{\cal O}^{J/\psi}({}^3S_1^{[1]})\rangle$ NRQCD matrix element.

For the Chao et al.\ LDME set at $\mu_\Lambda=m_c$, the summed two-photon CO contribution is $-0.0531~{\rm fb}$, corresponding to about $-13\%$ of the NLO two-photon CS result. Its magnitude is about $1.7$ times the previously calculated summed single-photon CO contribution. The result arises from a cancellation between the positive ${}^1S_0^{[8]}$ and negative ${}^3P_J^{[8]}$ contributions and is consequently sensitive to the ${}^3P_0^{[8]}$ LDME and the NRQCD factorization scale, while the charm-quark-mass dependence is relatively mild. 
At higher energies, the two-photon CS and CO contributions exhibit similar falloffs, with $n_{\rm eff}\simeq0.8$ over the energy range considered. Their different $J/\psi$ energy distributions further show that the relative importance of the two CO contributions varies across phase space.
Different LDME sets lead to substantial variations in both the magnitude and the sign of the summed CO contribution, which calls for more precise future experimental measurements to constrain the relevant LDMEs.

\section*{Acknowledgements} 
This work was supported by the National Natural Science Foundation of China (No. 11705078, 12575087).


\begin{thebibliography}{99}
	

	
	\bibitem{Bodwin:1994jh}
	G.~T. Bodwin, E.~Braaten, and G.~P. Lepage,
	Rigorous QCD analysis of inclusive annihilation and production of
	heavy quarkonium,
	Phys. Rev. D \textbf{51}, 1125 (1995);
	Erratum: Phys. Rev. D \textbf{55}, 5853 (1997).
	
	
		\bibitem{Braaten:1995CO}
	E.~Braaten and Y.-Q.~Chen,
	Signature for color-octet production of $J/\psi$ in
	$e^+e^-$ annihilation,
	Phys. Rev. Lett. \textbf{76}, 730 (1996).
	
	\bibitem{Butenschoen:2011yh}
	M.~Butensch\"on and B.~A. Kniehl,
	World data of $J/\psi$ production consolidate NRQCD factorization
	at NLO,
	Phys. Rev. D \textbf{84}, 051501 (2011).
	
	\bibitem{Butenschoen:2012px}
	M.~Butensch\"on and B.~A. Kniehl,
	$J/\psi$ polarization at Tevatron and LHC:
	Nonrelativistic-QCD factorization at the crossroads,
	Phys. Rev. Lett. \textbf{108}, 172002 (2012).
	
	\bibitem{Chao:2012iv}
	K. T.~Chao, Y.-Q.~Ma, H.-S.~Shao, K.~Wang, and Y.-J.~Zhang,
	$J/\psi$ polarization at hadron colliders in nonrelativistic QCD,
	Phys. Rev. Lett. \textbf{108}, 242004 (2012).
	
	\bibitem{Gong:2012ug}
	B.~Gong, L.-P.~Wan, J.-X.~Wang, and H.-F.~Zhang,
	Polarization for prompt $J/\psi$ and $\psi(2S)$ production
	at the Tevatron and LHC,
	Phys. Rev. Lett. \textbf{110}, 042002 (2013).
	
	\bibitem{Bodwin:2014gia}
	G.~T. Bodwin, H.~S. Chung, U.-R.~Kim, and J.~Lee,
	Fragmentation contributions to $J/\psi$ production at the
	Tevatron and the LHC,
	Phys. Rev. Lett. \textbf{113}, 022001 (2014).

	
	\bibitem{Zhang:2009CO}
	Y. J.~Zhang, Y.-Q.~Ma, K.~Wang, and K.-T.~Chao,
	QCD radiative correction to color-octet $J/\psi$ inclusive
	production at $B$ factories,
	Phys. Rev. D \textbf{81}, 034015 (2010).
	
	\bibitem{Chen:2022SGF}
	A.-P.~Chen, X.-B.~Jin, Y.-Q.~Ma, and C.~Meng,
	Color-octet contributions for $J/\psi$ inclusive production at
	$B$ factories in soft gluon factorization,
	JHEP \textbf{03}, 202 (2022).
	
	%\cite{Wang:2003fw}
	\bibitem{Wang:2003fw}
	J.~X.~Wang,
	The Color octet effect from $e^{+} e^{-} \to J/\psi$ + $X$ + $\gamma$ at $B$ factory,
	AIP Conf. Proc. \textbf{1092} (2009).
	%11 citations counted in INSPIRE as of 23 Nov 2026

	
	\bibitem{Bodwin:2002JpsiJpsi}
	G.~T. Bodwin, E.~Braaten, and J.~Lee,
	$e^+e^-$ annihilation into $J/\psi+J/\psi$,
	Phys. Rev. Lett. \textbf{90}, 162001 (2003).
	
	\bibitem{Bodwin:2002hg}
	G.~T. Bodwin, J.~Lee, and E.~Braaten,
	Exclusive double-charmonium production from $e^+e^-$ annihilation
	into two virtual photons,
	Phys. Rev. D \textbf{67}, 054023 (2003);
	Erratum: Phys. Rev. D \textbf{72}, 099904 (2005).
	
	\bibitem{Bodwin:2006ke}
	G.~T. Bodwin, E.~Braaten, J.~Lee, and C.~Yu,
	Exclusive two-vector-meson production from $e^+e^-$ annihilation,
	Phys. Rev. D \textbf{74}, 074014 (2006).
	
	\bibitem{Gong:2008ce}
	B.~Gong and J.-X.~Wang,
	QCD corrections to double $J/\psi$ production in $e^+e^-$
	annihilation at $\sqrt{s}=10.6~{\rm GeV}$,
	Phys. Rev. Lett. \textbf{100}, 181803 (2008).
	
	\bibitem{Fan:2012dy}
	Y.~Fan, J.~Lee, and C.~Yu,
	Resummation of relativistic corrections to exclusive productions of
	charmonia in $e^+e^-$ collisions,
	Phys. Rev. D \textbf{87}, 094032 (2013).
	
	\bibitem{Sang:2023double}
	W.-L.~Sang, F.~Feng, Y.~Jia, Z.~Mo, J.~Pan, and J.-Y.~Zhang,
	Optimized $\mathcal O(\alpha_s^2)$ correction to exclusive
	double-$J/\psi$ production at $B$ factories,
	Phys. Rev. Lett. \textbf{131}, 161904 (2023).
	
	\bibitem{Huang:2024double}
	X.-D.~Huang, B.~Gong, R.-C.~Niu, H.-M.~Yu, and J.-X.~Wang,
	Next-to-next-to-leading-order QCD corrections to double
	$J/\psi$ production at the $B$ factories,
	JHEP \textbf{02}, 055 (2024).
		
	%\cite{Liu:2026ray}
	\bibitem{Liu:2026ray}
	H.~Y.~Liu, C.~Li and W.~L.~Sang,
	Semi-analytical two-loop QCD corrections to e$^{+}$e$^{-}$ $\rightarrow$ J/$\psi$ + $\chi_{cJ}$ at B factories,
	JHEP \textbf{10} (2026), 004.

%\cite{Chang:2026vqf}
\bibitem{Chang:2026vqf}
X.~Chang, C.~Zeng, B.~Gong, J.~X.~Wang and C.~Li,
NNLO QCD corrections to direct $J/\psi$ production with open charm at B factories,
[arXiv:2610.39895 [hep-ph]].
%0 citations counted in INSPIRE as of 21 Nov 2026

	
	\bibitem{Belle:2004ab}
	K.~Abe \textit{et al.} (Belle Collaboration),
	Study of double charmonium production in $e^+e^-$ annihilation at
	$\sqrt{s}\simeq10.6~{\rm GeV}$,
	Phys. Rev. D \textbf{70}, 071102 (2004).
	
	\bibitem{BaBar:2005de}
	B.~Aubert \textit{et al.} (BABAR Collaboration),
	Measurement of double charmonium production in $e^+e^-$ annihilations
	at $\sqrt{s}=10.6~{\rm GeV}$,
	Phys. Rev. D \textbf{72}, 031101 (2005).
	

	\bibitem{Feng:2017}
	Y.~Feng, Z.~Sun, and H.-F.~Zhang,
	Is the colour-octet mechanism consistent with the double-$J/\psi$
	production measurement at $B$-factories?,
	Eur. Phys. J. C \textbf{77}, 221 (2017).


	\bibitem{Petrelli:1997ge}
	A.~Petrelli, M.~Cacciari, M.~Greco, F.~Maltoni, and M.~L. Mangano,
	NLO production and decay of quarkonium,
	Nucl. Phys. B \textbf{514}, 245 (1998).
	
	\bibitem{Harris:2001sx}
	B.~W. Harris and J.~F. Owens,
	The two cutoff phase space slicing method,
	Phys. Rev. D \textbf{65}, 094032 (2002).



   %\cite{Sun:2026hff}
   \bibitem{Sun:2026hff}
   Z.~Sun, S.~J.~Qi and Y.~Z.~Jiang,
   Inclusive productions of $J/\psi+\eta_c$ in $e^{+}e^{-}$ annihilation at Belle,
   [arXiv:2609.09590 [hep-ph]].
   %1 citations counted in INSPIRE as of 11 Oct 2026

	
	
%\cite{Wang:2026gso}
\bibitem{Wang:2026gso}
X.~P.~Wang, Y.~J.~Li, G.~Z.~Xu and K.~Y.~Liu,
Associated production of charmonia and bottomonia with color-octet channels at the Z factory, CEPC, and FCC-ee,
Phys. Rev. D \textbf{114} (2026) no.1, 016016


	
	\bibitem{Zhang:2014ybe}
	H.-F.~Zhang, Z.~Sun, W.-L.~Sang, and R.~Li,
	Impact of $\eta_c$ hadroproduction data on charmonium production
	and polarization within NRQCD framework,
	Phys. Rev. Lett. \textbf{114}, 092006 (2015).
	
	\bibitem{Sun:2015pia}
	Z.~Sun and H.-F.~Zhang,
	Reconciling charmonium production and polarization data within the
	nonrelativistic QCD framework,
	Chin. Phys. C \textbf{42}, 043104 (2018).

    \bibitem{Li:2014fya}
	Y.-J.~Li, G.-Z.~Xu, P.-P.~Zhang, Y.-J.~Zhang, and K.-Y.~Liu,
	Study of Color Octet Matrix Elements Through $J/\psi$ Production in $e^{+}e^{-}$ Annihilation,
	Eur. Phys. J. C \textbf{77}, 597 (2017), arXiv:1409.2293 [hep-ph].

	
\end{thebibliography}
\end{document}